\documentclass[aps,prfluids,showpacs,notitlepage,longbibliography]{revtex4-1}
\usepackage{epsfig,graphics,amssymb,amsmath,subeqnarray,color,bm,bbm}
\usepackage[toc,page]{appendix}
\usepackage{mathtools}
\usepackage{kpfonts}
\usepackage{xcolor}
\usepackage{float}
\usepackage{graphicx}
\usepackage[colorlinks=true,linkcolor=blue]{hyperref}

\begin{document}
\title{Dynamics of poroelastocapillary rise}
\author{Babak Nasouri}
\affiliation{Department of Mechanical Engineering, University of British Columbia, Vancouver, British Columbia, Canada V6T 1Z4}
\author{Benjamin Thorne}
\affiliation{Department of Mechanical Engineering, University of British Columbia, Vancouver, British Columbia, Canada V6T 1Z4}
\author{Gwynn J. Elfring}
\email{gelfring@mech.ubc.ca}
\affiliation{Department of Mechanical Engineering, University of British Columbia, Vancouver, British Columbia, Canada V6T 1Z4}
\date{\today}


\begin{abstract} 
A wetting liquid is driven through a thin gap due to surface tension and when the gap boundaries are elastic, the liquid deforms the gap as it rises. But when the fluid boundaries are also permeable (or \textit{poroelastic}), the liquid can permeate the boundaries as the fluid rises and change their properties, for example by swelling and softening, thereby altering the dynamics of the rise. In this paper, we study the dynamics of capillary rise between two poroelastic sheets to understand the effects of boundary permeability and softening. We find that if the bending rigidity of sheets is reduced, due to liquid permeation, the sheets coalesce faster compared to the case of impermeable sheets. We show that as a direct consequence of this faster coalescence, the volume of fluid captured between the sheets can be significantly lower.
\end{abstract}

\maketitle

\section{Introduction}

Interactions of capillary forces and elastic materials are abundant in nature:  bundle formations in bristles of a wet painting brush or in wet eyelashes \citep{bico2004,duprat2015book}, strong hydrophobic interactions of the feathers of aquatic birds \citep{rijke1970}, or the fluid-mediated adhesion of a beetle to a substrate \citep{eisner2000} are just a few examples amongst the many. These interactions, referred to as elastocapillary effects, have shown to be a key factor in collapsing  \citep{maboudian1997,cambau2011} or fabricating \citep{leong2010,crane2012} engineered microstructures, in the lubrication of soft materials \citep{hooke1972,snoeijer2016} and can be exploited for ultra-thin whitening \citep{chandra2009}. Our understanding of capillary rise dates back to experiments of Newton (1704), Jurin (1712), and the analysis of Laplace (1805). When a small tube is in contact with a wetting fluid, capillary forces drive the liquid into the tube until they are balanced by the gravitational forces. In a seminal work, \citet{washburn1921} showed that in capillary rise the balance of surface tension forces and viscous dissipation governs the rate of fluid motion.

The coupling of surface tension forces and elastic forces can lead to surface deformations when liquid is in contact with elastic media, for instance by forming wrinkles \citep{huang2007} or a ridge \citep{pericet2008}. In particular, capillary-induced self-assembly of thin flexible materials has attracted extensive attention, owing to recent developments in micro- and nano-engineering (see \citep{bico2018} and the references therein). To shed some light on such interactions, \citet{kim2006maha} notably studied the capillary rise of a liquid between two flexible sheets, clamped at one end and free at the other, and analytically characterized the equilibrium configurations. \citet{duprat2011} then complemented this analysis by looking into sheet deformation prior to equilibrium and developed a framework for capturing the dynamical behavior of this elastocapillary rise (see also \citep{aristoff2011}). Subsequently, several studies have further extended this model by, for instance, investigating its multiple equilibria \citep{taroni2012}, considering a series of sheets \citep{gat2013,singh2014} or by employing the model to enhance the capillary flow in microchannels \citep{anoop2015}. In all of these studies, the sheets have been taken as impermeable entities whose properties remain constant upon wetting. For example, paper sheets are often permeable as liquid may diffuse through and change their properties significantly \citep{lee2016}. For instance, water softens a paper napkin as it flows between the plies, which may affect its absorbency. Given that such absorbency is important in paper products used in household and diagnostic applications \citep{martinez2008}, here we try to quantify the effect of sheet permeability on elastocapillary rise, as a simple model of, for instance, flow between plies of a paper towel.

Paper consists of multiple layers of cellulose fibers. Each fiber has an internal cavity of half of its size, and the surrounding wall is closely packed with hydrophilic microfibrils \citep{topgaard2001}. When infiltrated by water, before filling the cavities, the liquid diffuses within the microfibrils and causes expansion and swelling \citep{enderby1955}. Imbibition of water into cellulose sponges \citep{kvick2017,ha2018}, swelling of two parallel sheets submerged into a liquid bath \citep{holmes2016}, and self-rolling of a piece of paper immersed into water \citep{reyssat2011} are all examples of this phenomenon. In a recent study, \citet{lee2016} characterized deformation and stiffness of a strip of a paper when it imbibes a stain of water due to surface tension forces. They showed that by absorbing water, the paper sheet swells, increasing its thickness by $\sim25\%$, while simultaneously decreasing its Young's modulus from $\hat{E}=828~\text{MPa}$ to $\hat{E}=24~\text{MPa}$. This significant change of stiffness, which occurs due to imbibition of water by fiber-based materials, can alter the papers absorptivity, and so is quite important in painting \citep{capodicasa2010} and diagnostic applications \citep{martinez2010}. Thus, to understand the effect of this permeability on capillary rise, in this paper, we consider the elastocapillary rise of a liquid (e.g., water) between two paper sheets. The paper sheets are permeable and they become softer as the liquid rises and permeates through. To study the system's behavior, we closely follow the work of \citet{duprat2011}, but modify their model to incorporate the permeability of the sheets by allowing the properties to change upon liquid imbibition. We discuss the dynamics of sheet deformation and the equilibrium states, and compare them with those of impermeable sheets. We show that due to the softening of the sheets, the absorbency of the system with permeable sheets is reduced compared to the system with impermeable ones.

The outline of this paper is as follows. In Sec.~\ref{7:setup}, we present the details of the system and perform a scaling analysis. In Sec.~\ref{7:equations}, we derive evolution equations for the sheet deformation and the meniscus position. We then solve these equations using a finite difference scheme given in Sec.~\ref{7:method}. Finally, in Sec.~\ref{7:results}, we discuss the results, compare them with those of impermeable sheets and our experimental observations.

\section{Problem Statement}
\label{7:setup}
\begin{figure}
\begin{center}
\includegraphics[width=0.85\textwidth]{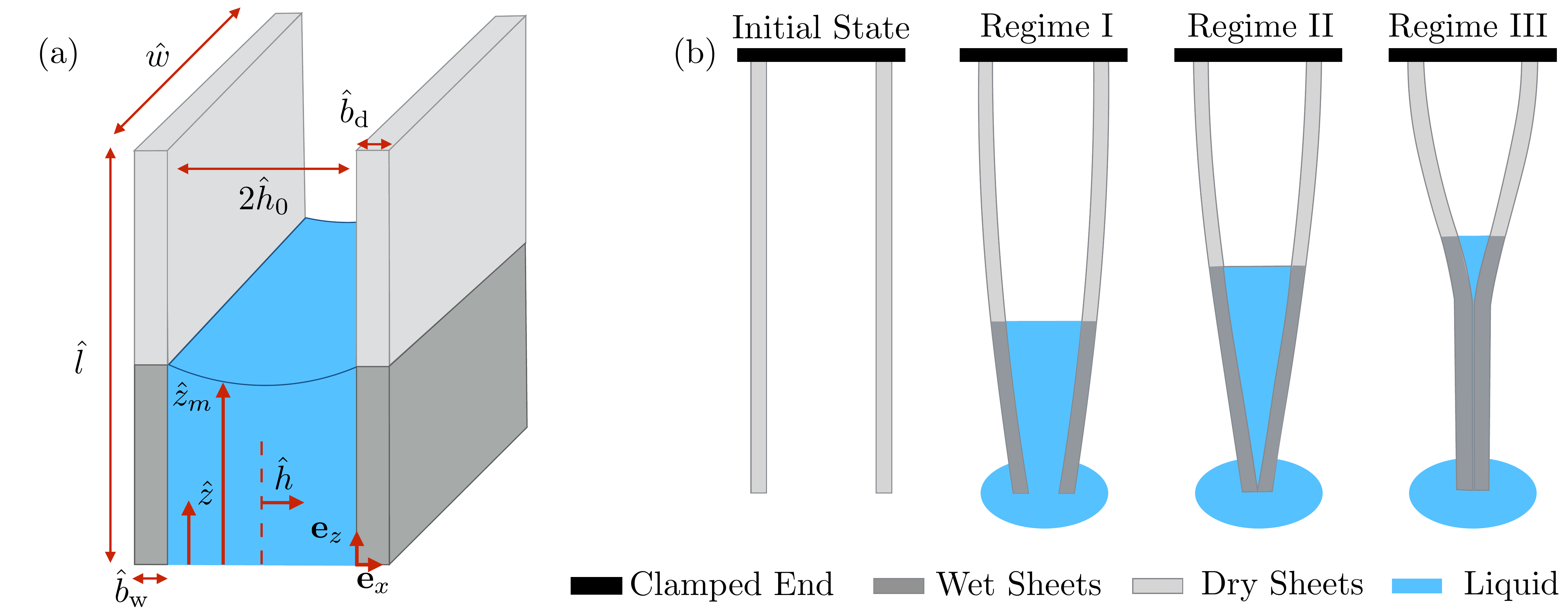}
\caption{The schematic of the system: (a) Two poroelastic sheets clamped at the upper end are immersed into a liquid bath from the lower end. (b) Three scenarios of the equilibrium. In regime I, sheets only slightly bend. In regime II, sheets lower end are in contact and in regime III, sheets coalesce over a finite length.}
\label{7:schematic}
\end{center}
\end{figure}

We are interested in the capillary rise of a viscous fluid between two permeable elastic paper sheets. We consider sheets of length $\hat{l}$, thickness $\hat{b}$, width $\hat{w}$ and assume they are separated initially by distance $2\hat{h}_0$, as shown in Fig.~\ref{7:schematic}(a). The sheets are clamped at the upper end (i.e., $\hat{z}=\hat{l}$) and immersed into the liquid bath from the lower free end (i.e., $\hat{z}=0$). The liquid then rises vertically (in $\mathbf{e}_z$) and we refer to the meniscus position as $\hat{z}_m$. The sheets are elastic and deform as the liquid rises and we quantify the deflections by  $\hat{h}(\hat{z},\hat{t})$ which varies from $\hat{h}_0$ (no deflection) to zero (coalescence). 
The behavior of the system is dominated by three forces: surface tension forces drive the flow, gravitational forces resist the rise of the liquid, and finally the elastic forces account for the sheet deformations. We characterize these forces using two dimensionless groups, namely the elastocapillary number $\mathcal{E}$, which compares surface tension forces to elastic forces, and the Bond number $\mathcal{B}$, which compares the effects of gravity and surface tension. We define
\begin{align}
\label{elastocapillary}
\mathcal{E}&=\frac{\hat{\gamma} \hat{l}^4}{\hat{B}\hat{h}^2_0},\\
\mathcal{B}&=\frac{\hat{\rho} \hat{g} \hat{l}\hat{h}_0}{\hat{\gamma}},
\end{align} 
where $\hat{\gamma}$ is the surface tension, $\hat{\rho}$ is density, $\hat{g}$ is the magnitude of gravitational acceleration, and $\hat{B}=\frac{1}{12}\hat{E} \hat{b}^3$ is the bending stiffness per unit width with $\hat{E}$ being the Young's modulus of the sheet. \citet{duprat2011} showed that depending on values of $\mathcal{B}$ and $\mathcal{E}$, the system can exhibit three different configurations: the sheets bend but do not touch ($\hat{h}(\hat{z}=0)>0$, regime I), they deflect such that they touch but do not coalesce ($\hat{h}(\hat{z}=0)=0$ and $\frac{\partial \hat{h}}{\partial \hat{z}}(\hat{z}=0)\neq0$, regime II), or they coalesce over a finite length (regime III), as shown in Fig.~\ref{7:schematic}(b). 

Since the sheets are permeable, as liquid rises due to surface tension forces, it also permeates through the sheets and changes their properties. Thus, to account for such changes, in our model we consider different properties for wet and dry parts of the sheets and henceforth refer to them using subscripts `\textit{w}' and `\textit{d}' as $\left\{\hat{B}_w, \mathcal{E}_w,\hat{b}_w, \cdots   \right\}$ and $\left\{\hat{B}_d, \mathcal{E}_d   ,\hat{b}_d,\cdots  \right\}$.

\subsection*{Scaling analysis}
This dynamical system involves several time scales and to simplify the problem it is important to compare them. The classic time scale over which fluid rises ($\hat{\tau}_r$), is found by employing a steady unidirectional (Poiseuille) flow field $u=\Delta \hat{p} \hat{h}_0^2/(3\hat{\mu} \hat{l})$ and noting that at equilibrium pressure must balance gravity $\Delta \hat{p} = \hat{\rho} \hat{g} \hat{l}$ and that the pressure is set by surface tension $\Delta \hat{p} = \hat{\gamma}/\hat{h}_0$ so that capillary rise time scale $\hat{\tau}_r = \hat{l}/\hat{u} = 3\hat{\mu} \hat{\gamma} /\hat{h}_0^3(\hat{\rho} \hat{g})^2$. When the elastic deformation of the sheets is appreciable, \citet{duprat2011} argued that the pressure scale is instead set by the deflection of the sheets $\Delta \hat{p} = \hat{B}\hat{h}_0/\hat{l}^4$ so that in this case the `visco-elastic' time scale $\tau_{ve} = \hat{l}/\hat{u} = 3\hat{\mu} \hat{l}^6/\hat{B} \hat{h}_0^3 = \mathcal{B}^2\mathcal{E}\hat{\tau}_r$. \citet{duprat2011} found that this occurs when $\mathcal{B}^2\mathcal{E}\gtrsim10$. Furthermore, at the moment when the sheets touch the liquid, inertia is clearly not negligible, and in fact dominates the capillary rise \citep{quere1997}. \citet{das2012} showed that this inertia-dominated regime prevails at $\hat{t}<\hat{\tau}_c$, where $\hat{\tau}_c=\sqrt{\hat{\rho} \hat{h}_0^3/{\hat{\gamma}}}$ (see also \citep{das2013}). In our problem fluid also permeates through thickness of the sheet as well as through the sheet vertically. Considering Washburn-like behavior for the fluid permeation in the sheet \citep{washburn1921} the time scale for lateral fluid permeation $\hat{\tau}_\text{D,$x$}={\hat{b}^2}/{\hat{D}_w}$ and vertical fluid permeation $\hat{\tau}_\text{D,$z$}={\hat{l}^2}/{\hat{D}_w}$, where $\hat{D}_w$ is the diffusion coefficient of the paper.

To compare these time scales, we use properties of a filter paper sheet reported in \citep{lee2016} and consider water as the viscous fluid. We also take $\hat{l}=1~$cm and $\hat{h}_0=0.5$ mm as typical values for the sheet length and gap size, respectively. We find $\hat{\tau}_c/\hat{\tau}_{r}\sim 10^{-2}$, so we may neglect the effect of inertia since it vanishes very fast. We also find $\hat{\tau}_\text{D,$x$}/\hat{\tau}_r\sim 10^{-2}$ indicating that fluid diffuses through the thickness of the sheet considerably faster than the capillary rise. Thus, we assume liquid permeates the sheet (in its thickness), instantly. We also find $\hat{\tau}_r/\hat{\tau}_\text{D,$z$}\sim 10^{-3}$, implying that fluid diffusion along the length of the sheet is very slow and so is negligible at the time scale of the rise. Relying on these scalings, one can assume that paper sheet is wet in the liquid-filled region and dry in the liquid-free region. Thus, given a meniscus position $\hat{z}_m$, we consider properties of the wet paper sheet for $0\leq\hat{z}\leq\hat{z}_m$ and dry one for $\hat{z}_m<\hat{z}\leq\hat{l}$.
\section{Governing Equations}
\label{7:equations}
Here, we take $\hat{l}$ as the characteristic vertical length scale, $\hat{h}_0$ as the characteristic deflection of the sheet, and $\hat{\tau}_{ve}$ as the characteristic time scale of the problem. We thereby non-dimensionalize all the quantities defining $z=\hat{z}/\hat{l}$, $z_m=\hat{z}_m/\hat{l}$, $h=\hat{h}/\hat{h}_0$, $t=\hat{t}/\hat{\tau}_{ve}$ and $p=\frac{\hat{p}}{\hat{B}_w\hat{h}_0/\hat{l}^4}$. Note that we have dropped the caret notation for the dimensionless parameters. We assume that a reflection symmetry between the sheets is maintained and so we derive the governing equations for one sheet (e.g., the one on the right in Fig.~\ref{7:schematic}(a)) and the other side shall be identical. In modeling the system, we closely follow the works of Stone et al. \citep{duprat2011,aristoff2011}, but incorporate the change of properties due to wetting. Since the properties of the sheet have a discontinuity at the meniscus, we treat the wet and dry parts separately and enforce boundary conditions at the interface. 

Noting that $\hat{h}_0\ll \hat{l}$, we employ the lubrication approximation to express the fluid motion. The one-dimensional momentum equation then dictates
\begin{align}
\label{7:ns}
&u=-h_w^2\left(\frac{\partial p}{\partial z}+ \mathcal{B}\mathcal{E}_w\right),
\end{align}
where $u$ is the vertical component of the meniscus velocity and all variables are averaged across the gap. Although the deflection of the sheet is a continuous function, we differentiate the deflection of the wet ($h_w$, when $z\le z_m$) and dry ($h_d$, when $z>z_m$) parts for clarity.
As noted earlier, the time scale of the liquid permeation through the sheet thickness is considerably smaller than the one of the rise. Thus, one can assume that the liquid saturates the sheets thickness instantly as the meniscus rises. Furthermore, because the sheet is very thin ($\hat{b}_w\ll\hat{h}_0$), we neglect the mass of liquid permeated within the sheet. One-dimentional mass conservation then yields
\begin{align}
\label{7:con}
&\frac{\partial h_w}{\partial t}+\frac{\partial}{\partial z} \left(h_w u\right)=0.
\end{align} 
Provided the sheet is sufficiently thin and long (i.e., $\hat{b}\ll\hat{w}\ll\hat{l}$), we may use linear elasticity to approximate the quasi-static sheet deflection
\begin{align}
\label{7:deflection}
p=\frac{\partial^4 h_w}{\partial z^4}.
\end{align}
Substituting pressure \eqref{7:deflection} to \eqref{7:ns}, we find
\begin{align}
\label{7:main2}
u=-h_w^2\left(\frac{\partial^5 h_w}{\partial z^5}+ \mathcal{B}\mathcal{E}_w\right),
\end{align}
which at $z=z_m$ gives the time evolution of the meniscus since $u(z=z_m)=\frac{\text{d} z_m}{\text{d} t}$. Now by making use of Eq. \eqref{7:main2}, we can rewrite the continuity equation \eqref{7:con} in terms of $h_w$ and its derivatives as
\begin{align}
\label{7:main}
\frac{\partial h_w}{\partial t}&=h_w^2\left( 3\frac{\partial h_w}{\partial z} \frac{\partial^5 h_w}{\partial z^5} +3\mathcal{B}\mathcal{E}_w \frac{\partial h_w}{\partial z}+ h_w\frac{\partial^6 h_w}{\partial z^6}  \right).
\end{align}
For $z>z_m$, there is no pressure gradient across the sheet and so we have
\begin{align}
\label{7:deflection2}
0=\frac{\partial^4 h_d}{\partial z^4}.
\end{align}

 \subsection*{Boundary conditions}

To account for the fixed end at $z=1$, we set $h_d(z=1)=1$ and $\frac{\partial h_d}{\partial z}(z=1)=0$. At the lower end ($z=0$), we note that there exists no net pressure or moment on the sheet so $\frac{\partial^4 h_w}{\partial z^4}(z=0)=\frac{\partial^2 h_w}{\partial z^2}(z=0)=0$. Furthermore, in regime I, wherein sheets cannot apply any force on each other $\frac{\partial^3 h_w}{\partial z^3}(z=0)=0$. In regime II, the sheets are no longer force free at $z=0$ but instead we have $h_w(z=0)=0$. For regime III, this boundary condition changes to $h_w(z=z_c)=0$ and $\frac{\partial h_w}{\partial z}(z=z_c)=0$, where $z_c$ is the length of the coalescence (i.e., $h_w(z<z_c)=0$).

At the interface $z=z_m$, continuity of sheet deflection and its slope are enforced by ${h_w(z=z_m)=h_d(z=z_m)}$ and ${\frac{\partial h_w}{\partial z}(z=z_m)=\frac{\partial h_d}{\partial z}(z=z_m)}$. Force and moment must also be continuous across the interface, thus $B_w\frac{\partial^3 h_w}{\partial z^3}=B_d \frac{\partial ^3h_d}{\partial z^3}$, and $B_w\frac{\partial^2 h_w}{\partial z^2}=B_d \frac{\partial ^2h_d}{\partial z^2}$. Finally, due to surface tension forces, there exists a pressure jump at the meniscus, which can be found using the Young-Laplace equation \citep{degennes2004} and can be written in dimensionless form as
\begin{equation}
\label{7:laplace}
\frac{\partial^4 {h_w}}{\partial z^4}-\frac{\partial^4 h_d}{\partial z^4}=-\frac{\mathcal{E}_w}{h_w},
\end{equation}
where we assume the contact angle is zero and neglect the effects of a dynamic contact angle.

In summary, we have a coupled system of sixth-order non-linear Partial Differential Equations (PDEs) subjected to two boundary conditions at the clamped end, three (or four in regime III) conditions at the free end and one condition accounting for the pressure jump at the meniscus. Additionally, we have four continuity conditions that need to be satisfied at the interface.

To solve this problem numerically, it is convenient to incorporate the dynamics of the liquid-free region into the boundary conditions for the liquid-filled region, given the simplicity of the governing equations in the liquid-free region \citep{duprat2011,aristoff2011}. To illustrate, from Eq.~\eqref{7:deflection2}, one can find the deflection and the slope at the liquid-free region as
\begin{align}
&h_d=A_3 z^3+A_2 z^2+A_1 z+A_0,\\
&\frac{\partial h_d}{\partial z}=A_6 z^2+A_5 z+A_4,
\end{align}
where $A_0$ to $A_6$ are determined using the boundary conditions at the fixed end ($z=1$) and the interface ($z=z_m$). $h_d$ and $\frac{\partial h_d}{\partial z}$ are then found in terms of $h_w$ and its derivatives at $z=z_m$ as
\begin{align}
\label{7:air}
h_d(z)&=1 +\frac{B_r}{6}(1-z)^2\left[3\frac{\partial^2 {h_w}}{\partial z^2} +\frac{\partial^3 {h_w}}{\partial z^3} (z-z_m) +2\frac{\partial^3 {h_w}}{\partial z^3}(1-z_m) \right],\\
\frac{\partial h_d}{\partial z}(z)&=-\frac{B_r}{2}(1-z)\left[2 \frac{\partial^2 {h_w}}{\partial z^2}+\frac{\partial^3{h_w}}{\partial z^3} (z-z_m) +\frac{\partial^3 {h_w}}{\partial z^3}(1-z_m)   \right],
\end{align}
where $B_r=\frac{B_w}{B_d}$. Noting that $h_w=h_d$ and $\frac{\partial {h_w}}{\partial z}=\frac{\partial h_d}{\partial z}$ at the interface, we then find
\begin{align}
\label{7:BC1}
&{h_w}(z=z_m)=1 +  \frac{B_r}{3}\frac{\partial^3 {h_w}}{\partial z^3} (1-z_m)^3 + \frac{B_r}{2}\frac{\partial^2 {h_w}}{\partial z^2} (1-z_m)^2, \\
\label{7:BC2}
&\frac{\partial {h_w}}{\partial z}(z=z_m)=-\frac{B_r}{2}\frac{\partial^3 {h_w}}{\partial z^3} (1-z_m)^2-B_r \frac{\partial^2 {h_w}}{\partial z^2} (1-z_m),
\end{align}
which provides two boundary conditions for $h_w$ at $z=z_m$. Now, the governing equations in the liquid-filled region are independent of $h_d$, and so we can determine the behavior of the sheet by solely solving the system for $z\leq z_m$. Once $h_w$ is found, we use Eq. \eqref{7:air} to find the deformation for the whole sheet. 
\section{Numerical Approach}
\label{7:method}
At early times, $t\ll 1$, one may consider a quasi-static deformation of the sheet ($\frac{\partial h_w}{\partial t}=0$) since sheet deflections are very small and time enters the problem only through boundary conditions \citep{duprat2011}. We expand the deflection $h_w(z)=\sum_{n=0}^{n=5} C_n z^n + {O}(z_m^6)$, where $z_m\ll1$. From the pressure drop at the meniscus given in \eqref{7:laplace}, we find $C_5=-\frac{\mathcal{E}_w}{120 z_m}$. Boundary conditions at $z=0$ dictate $C_4=C_2=0$ and, recalling that in this limit sheets can only reach regime I, we have $C_3=0$. Finally, from the boundary conditions given in \eqref{7:BC1} and \eqref{7:BC2}, we arrive at
\begin{align}
\label{7:asymp1}
h_w=1&-\frac{\mathcal{E}_w}{120z_m} z^5 + \left( \frac{\mathcal{E}_w z_m^3}{24}+\frac{\mathcal{E}_d z_m^3}{12} - \frac{\mathcal{E}_d z_m^2}{3} + \frac{\mathcal{E}_d z_m}{4}  \right)z\nonumber\\
&+\left( -\frac{\mathcal{E}_w z_m^3}{30}+\frac{\mathcal{E}_d z_m}{6} - \frac{\mathcal{E}_d }{6}  \right)z_m + {O}(z_m^6),
\end{align}
which governs the sheet deflection in the liquid-filled region for $t\ll 1$. Substituting $h_w$ from \eqref{7:asymp1} in \eqref{7:main2}, we find the leading-order evolution equation for the meniscus as 
\begin{align}
\label{7:asymp2}
\frac{\text{d} z_m}{\text{d} t}=\frac{\mathcal{E}_w}{z_m} -\mathcal{B} \mathcal{E}_w.
\end{align}
To solve the dynamical system in full, we use Eqs. \eqref{7:asymp1} and \eqref{7:asymp2} for early times, and then to determine the behavior of the system at later times, we follow the work of \citet{aristoff2011} and implement an implicit finite-difference scheme that is second-order accurate in space and first-order accurate in time. To resolve the nonlinear terms in Eqs. \eqref{7:main2} and \eqref{7:main} (e.g., $h_w^2\frac{\partial h_w}{\partial z} \frac{\partial^5 h_w}{\partial z^5}$), we discretize the higher-order term ($ \frac{\partial^5 h_w}{\partial z^5}$) and then use the results of the previous time-step for the lower-order terms ($h_w$ and $\frac{\partial h_w}{\partial z} $). We repeat the procedure iteratively until the relative convergence error reaches below $10^{-5}$. We discretize the sheet length in the liquid-filled region using 30 points and take $\Delta t=10^{-3}$ as the typical time step. Recalling that in the scale of the considered problem we found $\hat{\tau}_c/\hat{\tau}_{r}\sim 10^{-2}$,  we neglect inertia and take $h_w(t=0)=1$ (zero deflection) and $z_m(t=0)=10^{-3}$ (negligible inertial meniscus rise) as initial conditions. 

\section{Results and Discussion}
\label{7:results}
\begin{figure}
\begin{center}
\includegraphics[width=0.85\textwidth]{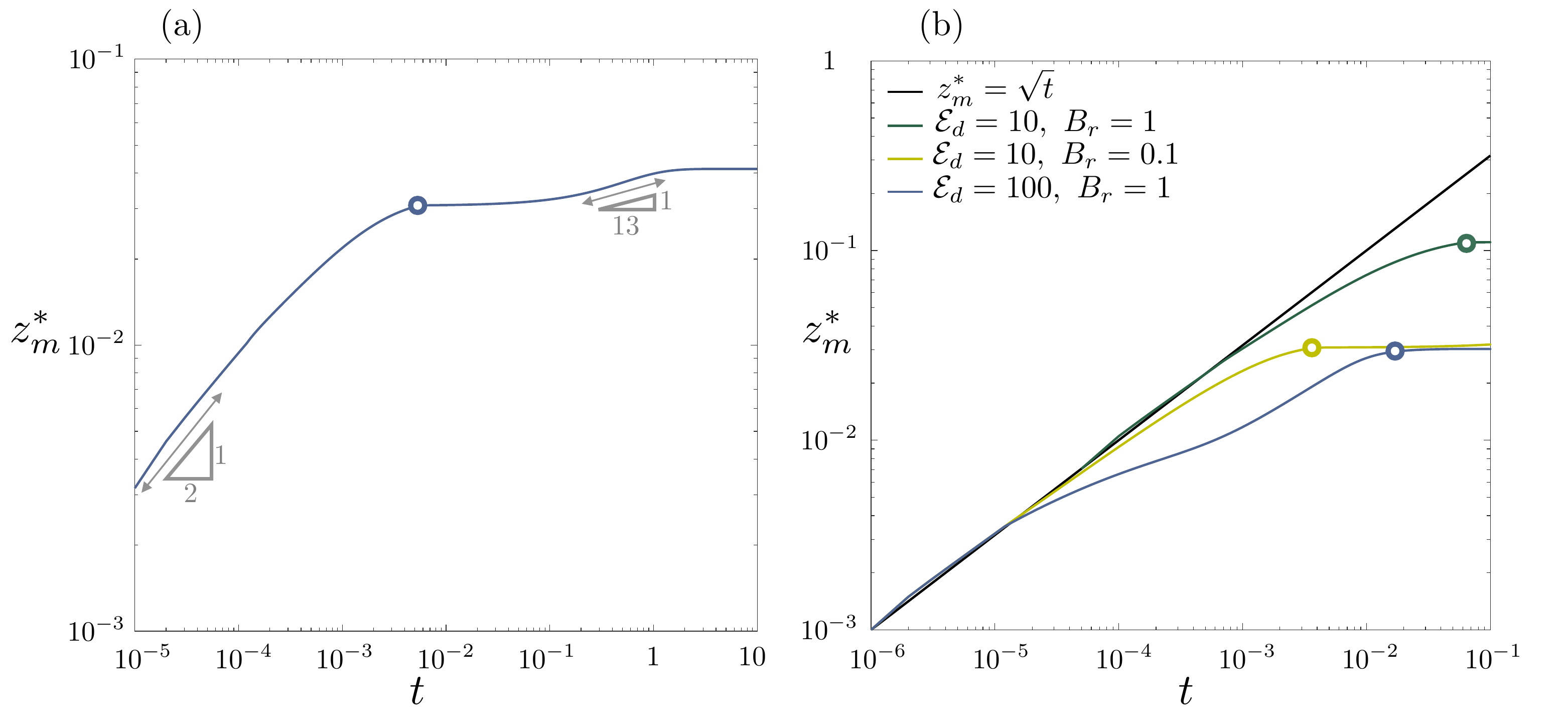}
\caption{The time evolution of rescaled meniscus, $z_m^*=\sqrt{{B_r}/({2\mathcal{E}_d)}}z_m$: (a) Permeable sheets with $\mathcal{B}=2$, $\mathcal{E}_d=10$ and $B_r=0.1$. (b) Two impermeable sheets with $\mathcal{E}_d=10$ and $\mathcal{E}_d=100$ and permeable sheets with $\mathcal{E}_d=10$ (or $\mathcal{E}_w=100$). In (b), for all cases  $\mathcal{B}=3$ and solid black line refers to the classical capillary rise \citep{washburn1921}. Circles on each line indicate the time in which sheets reach regime II ($h(t,z=0)=0$).}
\label{7:dyn}
\end{center}
\end{figure}

At early times ($t\ll1$) when the sheet deflection is not yet significant, elasticity does not contribute to the dynamics of the meniscus, nor does the permeability of sheet. Thus, regardless of the values of $\mathcal{E}_d$ and $B_r \left(=\frac{B_w}{B_d}=\frac{\mathcal{E}_d}{\mathcal{E}_w}\right)$, the meniscus strictly follows the classical behavior of a simple capillary rise in a rigid channel given by $z_m=\sqrt{2\mathcal{E}_w t}$ (note that in the dimensional form $\mathcal{E}_w$ disappears). Defining a rescaled meniscus position as $z_m^*=\sqrt{{1}/({2\mathcal{E}_w)}}z_m$ (or equivalently $z_m^*=\sqrt{{B_r}/({2\mathcal{E}_d)}}z_m$), one can see from Fig.~\ref{7:dyn}(a) that $z_m^*=\sqrt{t}$ predicts the initial behavior of $z_m^*$ quite accurately. This behavior can also be explained using the asymptotic expressions given in \eqref{7:asymp1} and \eqref{7:asymp2}. At leading order, we find $h_w=1$ indicating no deflection and so the problem is reduced to capillary rise between two rigid sheets. Furthermore, since at early times $z_m\ll 1$, meniscus dynamics can be approximated to leading order as ${\text{d} z_m}/{\text{d} t}={\mathcal{E}_w}/{z_m}$ (or ${\text{d} z_m^*}/{\text{d} t}={1}/{(2z_m^*)}$) confirming the diffusive behavior. At later times, the sheet deformation becomes appreciable and the meniscus position no longer follows the classical capillary-rise predictions. Once the lower ends of the sheets are in contact (denoted by a circle in Fig.~\ref{7:dyn}(a)), after a short period of almost stationary position, the meniscus rises with $z_m^*\sim t^{1/13}$, and then finally reaches the equilibrium.

\begin{figure}
\begin{center}
\includegraphics[scale=0.6]{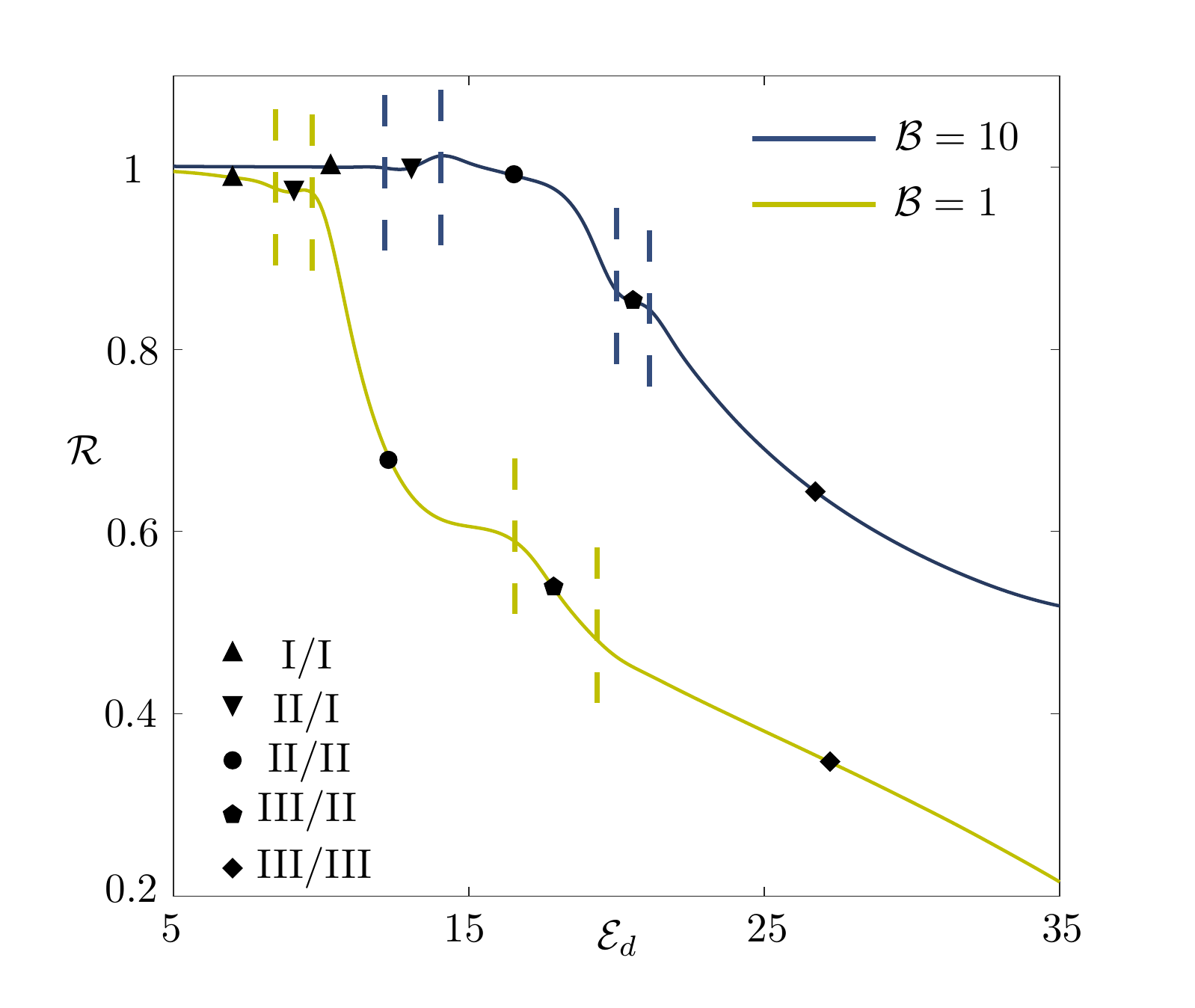}
\caption{The liquid absorption ratio, $\mathcal{R}$, of permeable sheets compared to impermeable ones of the same $\mathcal{E}_d$ for $\mathcal{B}=1$ and $\mathcal{B}=10$. Dashed lines separate the different combinations of equilibrium regimes for permeable and impermeable cases and symbols denote the regimes (e.g., II/I indicates that permeable sheets reach regime II while impermeable sheets are at regime I).}
\label{7:abs}
\end{center}
\end{figure}

We now compare the dynamics of the permeable sheets with impermeable cases. As noted earlier, for $t\ll 1$, one can neglect the effect of elasticity and permeability and so all the cases identically follow the classical behavior. For impermeable sheets, as the elastocapillary number increases, the effect of bending becomes more dominant and the meniscus position deviates from $z_m^*=\sqrt{t}$ sooner. But in permeable sheets, the time evolution of meniscus is dictated by elastocapillary numbers of both the dry region (e.g., $\mathcal{E}_d=10$) and the wet region (e.g., $\mathcal{E}_w=100$). Meniscus deviation from the classical behavior thereby lies within two cases of impermeable sheets with bounding elastocapillary numbers ($\mathcal{E}_d=10$ and $100$), as can be seen in Fig.~\ref{7:dyn}(b). However interestingly, the permeable sheets reach regime II ($h_w(z=0)=0$) faster than both bounding cases (denoted by circles in Fig.~\ref{7:dyn}(b)).

Note that the system can only collect liquid while $h_w(z=0)>0$; once the lower ends of sheets touch ($h_w(z=0)=0$), further rise of the liquid is purely due to sheet deflection as the liquid in the bath can no longer flow into the system. This behavior may indicate that permeable sheets have less time to capture the liquid. To better highlight this point, in Fig.~\ref{7:abs} we have reported the absorption ratio of permeable sheets to impermeable sheets, $\mathcal{R}$, by determining the area of the risen liquid between the sheets at equilibrium. As noted earlier, for small values of elastocapillary number, for which both cases reach regime I, the sheets only slightly bend. The effect of permeability is thereby not significant and the absorption ratio is nearly one. For permeable sheets, as the value of elastocapillary number increases, the sheets deflect more readily in response to the capillary rise. Thus, the softening of the sheets due to wetting facilitates this bending and leads the sheets to coalesce faster, thereby decreasing the absorption ratio. For instance, when $\mathcal{E}_d=35$, this ratio drops to $\sim40\%$ for $\mathcal{B}=10$ and $\sim 20\%$ for $\mathcal{B}=1$ in regime III.

\begin{figure}
\begin{center}
\includegraphics[scale=0.5]{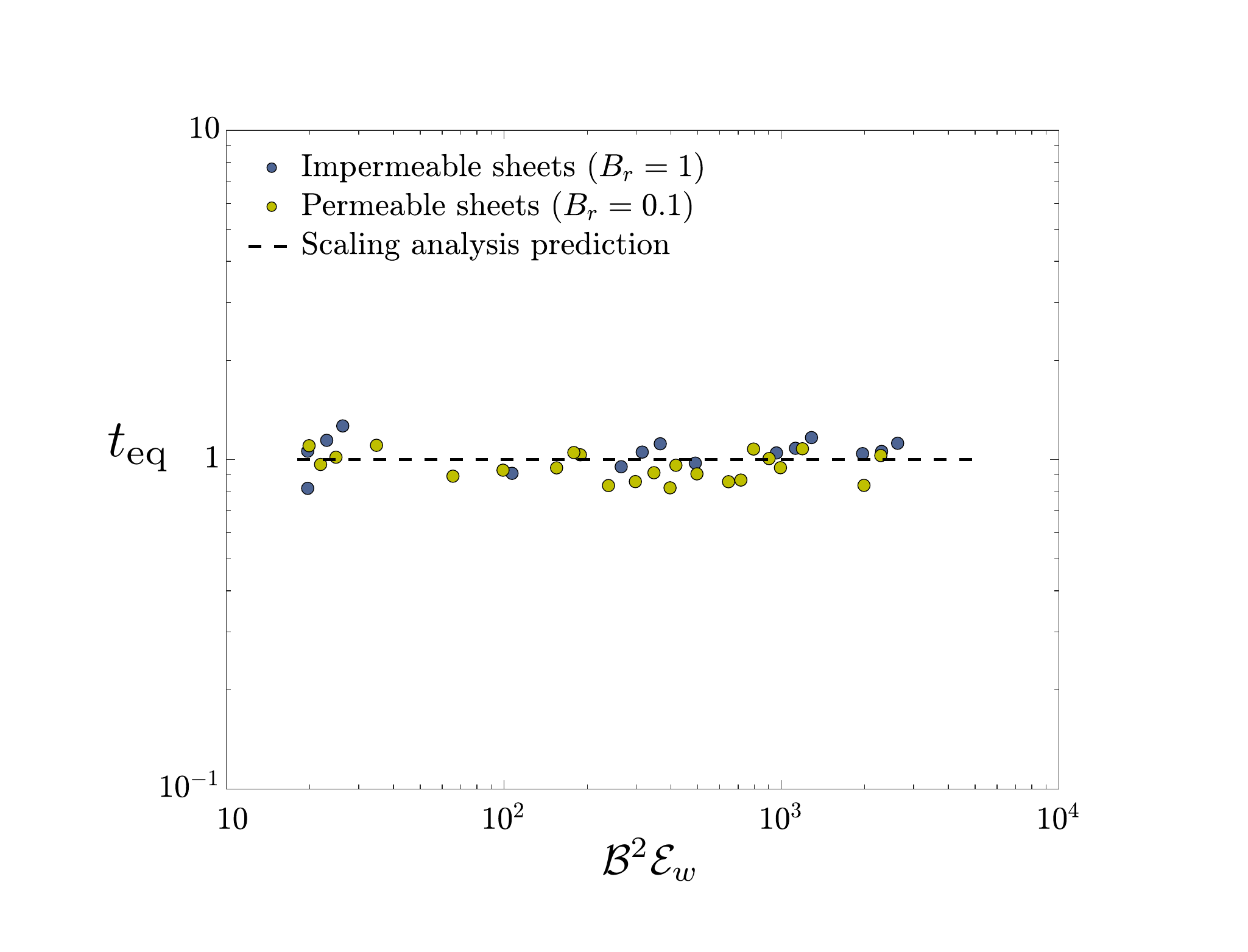}
\caption{Time to reach the 99\% of the equilibrium height versus $\mathcal{B}^2\mathcal{E}_w$. Symbols represent the numerical results for the equilibrium time scale of impermeable and permeable sheets. The dashed line indicates $t_\text{eq}=1$.}
\label{7:eq}
\end{center}
\end{figure}

Recall that \citet{duprat2011} found that for impermeable sheets when $\mathcal{B}^2\mathcal{E}_d\gtrsim10$ the time scale for the capillary rise is set by the deformation of the sheets, $\hat{\tau}_{ve}$. We find here, surprisingly, that despite the discontinuity of $\mathcal{E}$ at the meniscus, permeable sheets exhibit the same behavior for $\mathcal{B}^2\mathcal{E}_w\gtrsim10$, as shown in Fig.~\ref{7:eq} (note that in our dimensionless units $t=1$ is $\hat{t}=\hat{\tau}_{ve}$). This result indicates that when the sheets are sufficiently flexible, the equilibrium time scale for permeable sheets is dominantly dictated by the properties of the liquid-filled region, and the dry region has no appreciable contribution.

\begin{figure}[H]
\begin{center}
\includegraphics[width=0.85\textwidth]{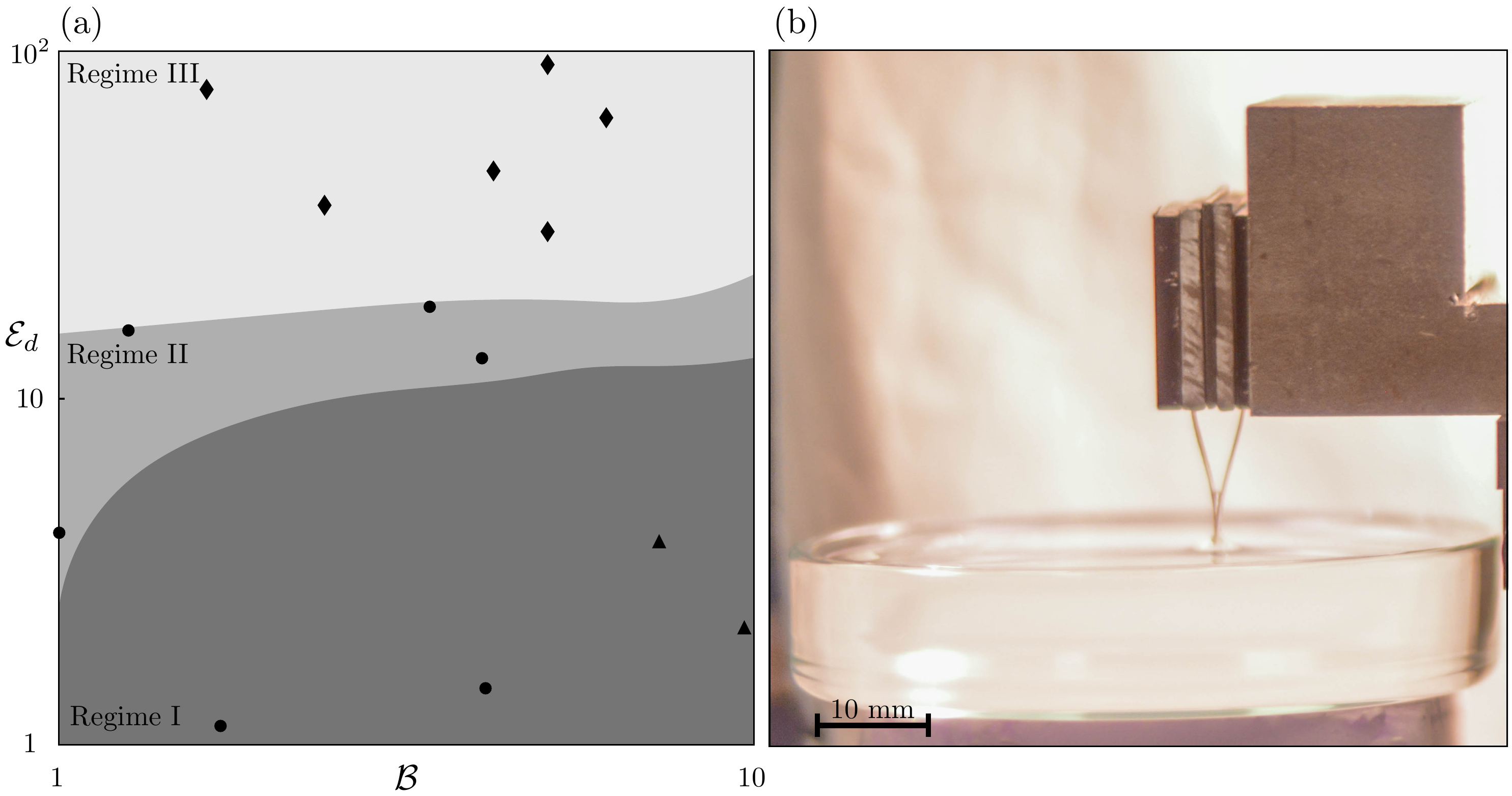}
\caption{(a) Regime map for the equilibrium sate of permeable sheets in a logarithmic $(\mathcal{B},\mathcal{E}_d)$ space. Each shade refers to its specified regime and is obtained using the numerical scheme discussed in Sec.~\ref{7:method}. Symbols are the equilibrium states observed experimentally: Triangles ($\blacktriangle$) denote regime I, circles ($\bullet$) indicate regime II and diamonds ($\blacklozenge$) refer to regime III. (b) The experimental apparatus. Sheets here are in regime III. }
\label{7:map}
\end{center}
\end{figure}

In Fig.~\ref{7:map}(a), the regime map for permeable sheets with $1<\mathcal{B}<10$ and $1<\mathcal{E}_d<10^2$ is illustrated. Unlike the case of impermeable sheets wherein for some values of elastocapillary number (i.e., $10 \lesssim\mathcal{E}_d \lesssim30$) regimes I and II coexist \citep{duprat2011}, here the three regimes are distinct, which may caused by further softening of the sheets due to wetting and their higher tendency to coalesce. Recall that higher Bond numbers can indicate larger gaps, so one can argue that sheets need to bend more to touch (regime II) or coalesce (regime III). Thus, reaching regime II and III is \textit{more difficult} when the Bond number ($\mathcal{B}$) is large. Indeed, as shown in Fig.~\ref{7:map}(a), our numerical results show that when the gap (or similarly $\mathcal{B}$) is large, reaching coalescence requires longer sheets which indicates higher values of elastocapillary number. We note that this is in contrast to impermeable sheets for which when the gap is large, coalescence is more easily obtained for shorter sheets \cite{duprat2011}.

Finally, in deriving the governing equations, we neglected the vertical liquid permeation within the sheet as it occurs on a longer time scale than the capillary rise. But, once the system reaches the equilibrium, the contribution of this upward permeation becomes significant, and the system then further evolves until the sheets are fully wet.

\subsection*{Experimental observations}
We compare our numerically-obtained regime map with experimental observations. We use filter papers (Whatmann grade 1) that we clamp at the upper end using rare-earth magnets, and set the initial gap using steel shims of varying thicknesses, as shown in Fig.~\ref{7:map}(b). We slowly immerse the sheets into a water bath and capture the equilibrium state using a digital camera. We report our observations for various combinations of Bond numbers and elastocapillary numbers in Fig.~\ref{7:map}(a) (symbols in the figure). We see that our numerical model is in good agreement with the experiments in predicting regime III. However, the model predictions and the experimental results deviate from each other at lower values of elastocapillary number for which the numerical model predicts regime I, while in the experiments, sheets surprisingly reach regime II. We believe that this discrepancy can arise from the inertial effects which were neglected in our model. To examine regime I, the system should have a low elastocapillary number. Noting that in our experiments we can only tune $\hat{h}_0$ and $\hat{l}$, we experimentally obtain low values for the elastocapillary number by increasing the initial gap between the sheets (see Eq. \eqref{elastocapillary}). But, the time scale of inertial effects $\hat{\tau}_c$ scales as $\sim\hat{h}_0^{3/2}$, and so it becomes more important when $\hat{h}_0$ increases. Specifically, for gap values of order $\sim2$ mm, the time scale of inertial effects is of the same order as that of the capillary rise. Therefore, for the cases with a large value of the initial gap (which often are in regime I), the effects of inertia is no longer negligible and may contribute to the equilibrium state. In our experiments, we observe that in this regime sheets rapidly (and significantly) bend toward each other at the very beginning and then exhibit the expected elastocapillary rise dynamics thereafter. Due to this initial effect, sheets' lower ends meet more readily and so the system proceeds to regime II.

\section{Conclusion}
In this paper, we studied the dynamics of capillary rise of a liquid between two flexible permeable paper sheets. Accounting for the change of sheet stiffness due to wetting, we discussed the motion of the meniscus and the sheet deflection as the system reaches the equilibrium. As the liquid rises, it permeates within the sheets and softens them significantly. Noting that the dynamics of the system is governed by both dry and wet regions of the sheets, as a direct consequence of this further softening in the wet region, permeable sheets evolve toward coalescence more readily, compared to impermeable ones. We also showed that the equilibrium time scale of the system is quite similar to those of impermeable sheets, however, the volume of fluid captured between the permeable sheets can be notably lower. The lower ends of permeable sheets meet sooner which means the system has less time to draw liquid in from the bath, and also the sheets are softer and so they deform significantly in response to the liquid rise. Recalling that, for instance, multi-ply paper towel is a collection of compressed multi-layer permeable sheets, our results indicate that permeability of fibers should be accounted for in modeling capillary-based systems, such as in sorption of oil spills \citep{nguyen2013,gui2013} and in designing microfluidic paper-based analytical devices \citep{martinez2008,martinez2010}. Given the generality of the considered model, our results can be extended to capture the behavior of a series of permeable sheets \citep{gat2013,singh2014} and be adapted to study the buckling of papers when they are fixed at both ends.

\section*{Acknowledgments}
The authors acknowledge funding from the NSERC Grant EGP 514715-2017.

\bibliography{reference}

\begin{thebibliography}{38}%
\makeatletter
\providecommand \@ifxundefined [1]{%
 \@ifx{#1\undefined}
}%
\providecommand \@ifnum [1]{%
 \ifnum #1\expandafter \@firstoftwo
 \else \expandafter \@secondoftwo
 \fi
}%
\providecommand \@ifx [1]{%
 \ifx #1\expandafter \@firstoftwo
 \else \expandafter \@secondoftwo
 \fi
}%
\providecommand \natexlab [1]{#1}%
\providecommand \enquote  [1]{``#1''}%
\providecommand \bibnamefont  [1]{#1}%
\providecommand \bibfnamefont [1]{#1}%
\providecommand \citenamefont [1]{#1}%
\providecommand \href@noop [0]{\@secondoftwo}%
\providecommand \href [0]{\begingroup \@sanitize@url \@href}%
\providecommand \@href[1]{\@@startlink{#1}\@@href}%
\providecommand \@@href[1]{\endgroup#1\@@endlink}%
\providecommand \@sanitize@url [0]{\catcode `\\12\catcode `\$12\catcode
  `\&12\catcode `\#12\catcode `\^12\catcode `\_12\catcode `\%12\relax}%
\providecommand \@@startlink[1]{}%
\providecommand \@@endlink[0]{}%
\providecommand \url  [0]{\begingroup\@sanitize@url \@url }%
\providecommand \@url [1]{\endgroup\@href {#1}{\urlprefix }}%
\providecommand \urlprefix  [0]{URL }%
\providecommand \Eprint [0]{\href }%
\providecommand \doibase [0]{http://dx.doi.org/}%
\providecommand \selectlanguage [0]{\@gobble}%
\providecommand \bibinfo  [0]{\@secondoftwo}%
\providecommand \bibfield  [0]{\@secondoftwo}%
\providecommand \translation [1]{[#1]}%
\providecommand \BibitemOpen [0]{}%
\providecommand \bibitemStop [0]{}%
\providecommand \bibitemNoStop [0]{.\EOS\space}%
\providecommand \EOS [0]{\spacefactor3000\relax}%
\providecommand \BibitemShut  [1]{\csname bibitem#1\endcsname}%
\let\auto@bib@innerbib\@empty
\bibitem [{\citenamefont {Bico}\ \emph {et~al.}(2004)\citenamefont {Bico},
  \citenamefont {Roman}, \citenamefont {Moulin},\ and\ \citenamefont
  {Boudaoud}}]{bico2004}%
  \BibitemOpen
  \bibfield  {author} {\bibinfo {author} {\bibfnamefont {J.}~\bibnamefont
  {Bico}}, \bibinfo {author} {\bibfnamefont {B.}~\bibnamefont {Roman}},
  \bibinfo {author} {\bibfnamefont {L.}~\bibnamefont {Moulin}}, \ and\ \bibinfo
  {author} {\bibfnamefont {A.}~\bibnamefont {Boudaoud}},\ }\bibfield  {title}
  {\enquote {\bibinfo {title} {Adhesion: Elastocapillary coalescence in wet
  hair},}\ }\href {\doibase 10.1038/432690a} {\bibfield  {journal} {\bibinfo
  {journal} {Nature}\ }\textbf {\bibinfo {volume} {432}},\ \bibinfo {pages}
  {690--690} (\bibinfo {year} {2004})}\BibitemShut {NoStop}%
\bibitem [{\citenamefont {Duprat}\ and\ \citenamefont
  {Stone}(2015)}]{duprat2015book}%
  \BibitemOpen
  \bibinfo {editor} {\bibfnamefont {C.}~\bibnamefont {Duprat}}\ and\ \bibinfo
  {editor} {\bibfnamefont {H.~A.}\ \bibnamefont {Stone}},\ eds.,\ \href
  {\doibase 10.1039/9781782628491} {\emph {\bibinfo {title} {Fluid-Structure
  Interactions in Low-{R}eynolds-Number Flows}}}\ (\bibinfo  {publisher} {Royal
  Society of Chemistry},\ \bibinfo {year} {2015})\BibitemShut {NoStop}%
\bibitem [{\citenamefont {Rijke}(1970)}]{rijke1970}%
  \BibitemOpen
  \bibfield  {author} {\bibinfo {author} {\bibfnamefont {A.~M.}\ \bibnamefont
  {Rijke}},\ }\bibfield  {title} {\enquote {\bibinfo {title} {Wettability and
  phylogenetic development of feather structure in water birds},}\ }\href@noop
  {} {\bibfield  {journal} {\bibinfo  {journal} {J. Exp. Biol.}\ }\textbf
  {\bibinfo {volume} {52}},\ \bibinfo {pages} {469--479} (\bibinfo {year}
  {1970})}\BibitemShut {NoStop}%
\bibitem [{\citenamefont {Eisner}\ and\ \citenamefont
  {Aneshansley}(2000)}]{eisner2000}%
  \BibitemOpen
  \bibfield  {author} {\bibinfo {author} {\bibfnamefont {T.}~\bibnamefont
  {Eisner}}\ and\ \bibinfo {author} {\bibfnamefont {D.~J.}\ \bibnamefont
  {Aneshansley}},\ }\bibfield  {title} {\enquote {\bibinfo {title} {Defense by
  foot adhesion in a beetle (hemisphaerota cyanea)},}\ }\href {\doibase
  10.1073/pnas.97.12.6568} {\bibfield  {journal} {\bibinfo  {journal} {Proc.
  Natl. Acad. Sci.}\ }\textbf {\bibinfo {volume} {97}},\ \bibinfo {pages}
  {6568--6573} (\bibinfo {year} {2000})}\BibitemShut {NoStop}%
\bibitem [{\citenamefont {Maboudian}(1997)}]{maboudian1997}%
  \BibitemOpen
  \bibfield  {author} {\bibinfo {author} {\bibfnamefont {R.}~\bibnamefont
  {Maboudian}},\ }\bibfield  {title} {\enquote {\bibinfo {title} {Critical
  review: Adhesion in surface micromechanical structures},}\ }\href {\doibase
  10.1116/1.589247} {\bibfield  {journal} {\bibinfo  {journal} {J. Vac. Sci.
  Technol. B Microelectron. Nanometer. Struct. Process. Meas. Phenom.}\
  }\textbf {\bibinfo {volume} {15}},\ \bibinfo {pages} {1} (\bibinfo {year}
  {1997})}\BibitemShut {NoStop}%
\bibitem [{\citenamefont {Cambau}\ \emph {et~al.}(2011)\citenamefont {Cambau},
  \citenamefont {Bico},\ and\ \citenamefont {Reyssat}}]{cambau2011}%
  \BibitemOpen
  \bibfield  {author} {\bibinfo {author} {\bibfnamefont {T.}~\bibnamefont
  {Cambau}}, \bibinfo {author} {\bibfnamefont {J.}~\bibnamefont {Bico}}, \ and\
  \bibinfo {author} {\bibfnamefont {E.}~\bibnamefont {Reyssat}},\ }\bibfield
  {title} {\enquote {\bibinfo {title} {Capillary rise between flexible
  walls},}\ }\href {\doibase 10.1209/0295-5075/96/24001} {\bibfield  {journal}
  {\bibinfo  {journal} {Europhys. Lett.}\ }\textbf {\bibinfo {volume} {96}},\
  \bibinfo {pages} {24001} (\bibinfo {year} {2011})}\BibitemShut {NoStop}%
\bibitem [{\citenamefont {Leong}\ \emph {et~al.}(2010)\citenamefont {Leong},
  \citenamefont {Zarafshar},\ and\ \citenamefont {Gracias}}]{leong2010}%
  \BibitemOpen
  \bibfield  {author} {\bibinfo {author} {\bibfnamefont {T.~G.}\ \bibnamefont
  {Leong}}, \bibinfo {author} {\bibfnamefont {A.~M.}\ \bibnamefont
  {Zarafshar}}, \ and\ \bibinfo {author} {\bibfnamefont {D.~H.}\ \bibnamefont
  {Gracias}},\ }\bibfield  {title} {\enquote {\bibinfo {title}
  {Three-€dimensional fabrication at small size scales},}\ }\href {\doibase
  10.1002/smll.200901704} {\bibfield  {journal} {\bibinfo  {journal} {Small}\
  }\textbf {\bibinfo {volume} {6}},\ \bibinfo {pages} {792--806} (\bibinfo
  {year} {2010})}\BibitemShut {NoStop}%
\bibitem [{\citenamefont {Crane}\ \emph {et~al.}(2012)\citenamefont {Crane},
  \citenamefont {Onen}, \citenamefont {Carballo}, \citenamefont {Ni},\ and\
  \citenamefont {Guldiken}}]{crane2012}%
  \BibitemOpen
  \bibfield  {author} {\bibinfo {author} {\bibfnamefont {N.~B.}\ \bibnamefont
  {Crane}}, \bibinfo {author} {\bibfnamefont {O.}~\bibnamefont {Onen}},
  \bibinfo {author} {\bibfnamefont {J.}~\bibnamefont {Carballo}}, \bibinfo
  {author} {\bibfnamefont {Q.}~\bibnamefont {Ni}}, \ and\ \bibinfo {author}
  {\bibfnamefont {R.}~\bibnamefont {Guldiken}},\ }\bibfield  {title} {\enquote
  {\bibinfo {title} {Fluidic assembly at the microscale: progress and
  prospects},}\ }\href {\doibase 10.1007/s10404-012-1060-1} {\bibfield
  {journal} {\bibinfo  {journal} {Microfluid. Nanofluid.}\ }\textbf {\bibinfo
  {volume} {14}},\ \bibinfo {pages} {383--419} (\bibinfo {year}
  {2012})}\BibitemShut {NoStop}%
\bibitem [{\citenamefont {Hooke}\ and\ \citenamefont
  {O'Donoghue}(1972)}]{hooke1972}%
  \BibitemOpen
  \bibfield  {author} {\bibinfo {author} {\bibfnamefont {C.~J.}\ \bibnamefont
  {Hooke}}\ and\ \bibinfo {author} {\bibfnamefont {J.~P.}\ \bibnamefont
  {O'Donoghue}},\ }\bibfield  {title} {\enquote {\bibinfo {title}
  {Elastohydrodynamic lubrication of soft, highly deformed contacts},}\ }\href
  {\doibase 10.1243/jmes_jour_1972_014_008_02} {\bibfield  {journal} {\bibinfo
  {journal} {J. Mech. Eng. Sci.}\ }\textbf {\bibinfo {volume} {14}},\ \bibinfo
  {pages} {34--48} (\bibinfo {year} {1972})}\BibitemShut {NoStop}%
\bibitem [{\citenamefont {Snoeijer}(2016)}]{snoeijer2016}%
  \BibitemOpen
  \bibfield  {author} {\bibinfo {author} {\bibfnamefont {J.~H.}\ \bibnamefont
  {Snoeijer}},\ }\bibfield  {title} {\enquote {\bibinfo {title} {Analogies
  between elastic and capillary interfaces},}\ }\href {\doibase
  10.1103/physrevfluids.1.060506} {\bibfield  {journal} {\bibinfo  {journal}
  {Phys. Rev. Fluids}\ }\textbf {\bibinfo {volume} {1}},\ \bibinfo {pages}
  {060506} (\bibinfo {year} {2016})}\BibitemShut {NoStop}%
\bibitem [{\citenamefont {Chandra}\ \emph {et~al.}(2009)\citenamefont
  {Chandra}, \citenamefont {Yang}, \citenamefont {Soshinsky},\ and\
  \citenamefont {Gambogi}}]{chandra2009}%
  \BibitemOpen
  \bibfield  {author} {\bibinfo {author} {\bibfnamefont {D.}~\bibnamefont
  {Chandra}}, \bibinfo {author} {\bibfnamefont {S.}~\bibnamefont {Yang}},
  \bibinfo {author} {\bibfnamefont {A.~A.}\ \bibnamefont {Soshinsky}}, \ and\
  \bibinfo {author} {\bibfnamefont {R.~J.}\ \bibnamefont {Gambogi}},\
  }\bibfield  {title} {\enquote {\bibinfo {title} {Biomimetic ultrathin
  whitening by capillary-force-induced random clustering of hydrogel
  micropillar arrays},}\ }\href {\doibase 10.1021/am900253z} {\bibfield
  {journal} {\bibinfo  {journal} {{ACS} Appl. Mater. Interfaces}\ }\textbf
  {\bibinfo {volume} {1}},\ \bibinfo {pages} {1698--1704} (\bibinfo {year}
  {2009})}\BibitemShut {NoStop}%
\bibitem [{\citenamefont {Washburn}(1921)}]{washburn1921}%
  \BibitemOpen
  \bibfield  {author} {\bibinfo {author} {\bibfnamefont {E.~W.}\ \bibnamefont
  {Washburn}},\ }\bibfield  {title} {\enquote {\bibinfo {title} {The dynamics
  of capillary flow},}\ }\href {\doibase 10.1103/physrev.17.273} {\bibfield
  {journal} {\bibinfo  {journal} {Phys. Rev.}\ }\textbf {\bibinfo {volume}
  {17}},\ \bibinfo {pages} {273--283} (\bibinfo {year} {1921})}\BibitemShut
  {NoStop}%
\bibitem [{\citenamefont {Huang}\ \emph {et~al.}(2007)\citenamefont {Huang},
  \citenamefont {Juszkiewicz}, \citenamefont {de~Jeu}, \citenamefont {Cerda},
  \citenamefont {Emrick}, \citenamefont {Menon},\ and\ \citenamefont
  {Russell}}]{huang2007}%
  \BibitemOpen
  \bibfield  {author} {\bibinfo {author} {\bibfnamefont {J.}~\bibnamefont
  {Huang}}, \bibinfo {author} {\bibfnamefont {M.}~\bibnamefont {Juszkiewicz}},
  \bibinfo {author} {\bibfnamefont {W.~H.}\ \bibnamefont {de~Jeu}}, \bibinfo
  {author} {\bibfnamefont {E.}~\bibnamefont {Cerda}}, \bibinfo {author}
  {\bibfnamefont {T.}~\bibnamefont {Emrick}}, \bibinfo {author} {\bibfnamefont
  {N.}~\bibnamefont {Menon}}, \ and\ \bibinfo {author} {\bibfnamefont {T.~P.}\
  \bibnamefont {Russell}},\ }\bibfield  {title} {\enquote {\bibinfo {title}
  {Capillary wrinkling of floating thin polymer films},}\ }\href {\doibase
  10.1126/science.1144616} {\bibfield  {journal} {\bibinfo  {journal}
  {Science}\ }\textbf {\bibinfo {volume} {317}},\ \bibinfo {pages} {650--653}
  (\bibinfo {year} {2007})}\BibitemShut {NoStop}%
\bibitem [{\citenamefont {Pericet-Camara}\ \emph {et~al.}(2008)\citenamefont
  {Pericet-Camara}, \citenamefont {Best}, \citenamefont {Butt},\ and\
  \citenamefont {Bonaccurso}}]{pericet2008}%
  \BibitemOpen
  \bibfield  {author} {\bibinfo {author} {\bibfnamefont {R.}~\bibnamefont
  {Pericet-Camara}}, \bibinfo {author} {\bibfnamefont {A.}~\bibnamefont
  {Best}}, \bibinfo {author} {\bibfnamefont {H.}~\bibnamefont {Butt}}, \ and\
  \bibinfo {author} {\bibfnamefont {E.}~\bibnamefont {Bonaccurso}},\ }\bibfield
   {title} {\enquote {\bibinfo {title} {Effect of capillary pressure and
  surface tension on the deformation of elastic surfaces by sessile liquid
  microdrops: An experimental investigation},}\ }\href {\doibase
  10.1021/la801862m} {\bibfield  {journal} {\bibinfo  {journal} {Langmuir}\
  }\textbf {\bibinfo {volume} {24}},\ \bibinfo {pages} {10565--10568} (\bibinfo
  {year} {2008})}\BibitemShut {NoStop}%
\bibitem [{\citenamefont {Bico}\ \emph {et~al.}(2018)\citenamefont {Bico},
  \citenamefont {Reyssat},\ and\ \citenamefont {Roman}}]{bico2018}%
  \BibitemOpen
  \bibfield  {author} {\bibinfo {author} {\bibfnamefont {J.}~\bibnamefont
  {Bico}}, \bibinfo {author} {\bibfnamefont {{\'{E}}.}~\bibnamefont {Reyssat}},
  \ and\ \bibinfo {author} {\bibfnamefont {B.}~\bibnamefont {Roman}},\
  }\bibfield  {title} {\enquote {\bibinfo {title} {Elastocapillarity: When
  surface tension deforms elastic solids},}\ }\href {\doibase
  10.1146/annurev-fluid-122316-050130} {\bibfield  {journal} {\bibinfo
  {journal} {Annu. Rev. Fluid Mech}\ }\textbf {\bibinfo {volume} {50}},\
  \bibinfo {pages} {629--659} (\bibinfo {year} {2018})}\BibitemShut {NoStop}%
\bibitem [{\citenamefont {Kim}\ and\ \citenamefont
  {Mahadevan}(2006)}]{kim2006maha}%
  \BibitemOpen
  \bibfield  {author} {\bibinfo {author} {\bibfnamefont {H.}~\bibnamefont
  {Kim}}\ and\ \bibinfo {author} {\bibfnamefont {L.}~\bibnamefont
  {Mahadevan}},\ }\bibfield  {title} {\enquote {\bibinfo {title} {Capillary
  rise between elastic sheets},}\ }\href {\doibase 10.1017/s0022112005007718}
  {\bibfield  {journal} {\bibinfo  {journal} {J. Fluid Mech.}\ }\textbf
  {\bibinfo {volume} {548}},\ \bibinfo {pages} {141} (\bibinfo {year}
  {2006})}\BibitemShut {NoStop}%
\bibitem [{\citenamefont {Duprat}\ \emph {et~al.}(2011)\citenamefont {Duprat},
  \citenamefont {Aristoff},\ and\ \citenamefont {Stone}}]{duprat2011}%
  \BibitemOpen
  \bibfield  {author} {\bibinfo {author} {\bibfnamefont {C.}~\bibnamefont
  {Duprat}}, \bibinfo {author} {\bibfnamefont {J.~M.}\ \bibnamefont
  {Aristoff}}, \ and\ \bibinfo {author} {\bibfnamefont {H.~A.}\ \bibnamefont
  {Stone}},\ }\bibfield  {title} {\enquote {\bibinfo {title} {Dynamics of
  elastocapillary rise},}\ }\href {\doibase 10.1017/jfm.2011.173} {\bibfield
  {journal} {\bibinfo  {journal} {J. Fluid. Mech.}\ }\textbf {\bibinfo {volume}
  {679}},\ \bibinfo {pages} {641--654} (\bibinfo {year} {2011})}\BibitemShut
  {NoStop}%
\bibitem [{\citenamefont {Aristoff}\ \emph {et~al.}(2011)\citenamefont
  {Aristoff}, \citenamefont {Duprat},\ and\ \citenamefont
  {Stone}}]{aristoff2011}%
  \BibitemOpen
  \bibfield  {author} {\bibinfo {author} {\bibfnamefont {J.~M.}\ \bibnamefont
  {Aristoff}}, \bibinfo {author} {\bibfnamefont {C.}~\bibnamefont {Duprat}}, \
  and\ \bibinfo {author} {\bibfnamefont {H.~A.}\ \bibnamefont {Stone}},\
  }\bibfield  {title} {\enquote {\bibinfo {title} {Elastocapillary
  imbibition},}\ }\href {\doibase 10.1016/j.ijnonlinmec.2010.09.001} {\bibfield
   {journal} {\bibinfo  {journal} {Int. J. Nonlinear Mech.}\ }\textbf {\bibinfo
  {volume} {46}},\ \bibinfo {pages} {648--656} (\bibinfo {year}
  {2011})}\BibitemShut {NoStop}%
\bibitem [{\citenamefont {Taroni}\ and\ \citenamefont
  {Vella}(2012)}]{taroni2012}%
  \BibitemOpen
  \bibfield  {author} {\bibinfo {author} {\bibfnamefont {M.}~\bibnamefont
  {Taroni}}\ and\ \bibinfo {author} {\bibfnamefont {D.}~\bibnamefont {Vella}},\
  }\bibfield  {title} {\enquote {\bibinfo {title} {Multiple equilibria in a
  simple elastocapillary system},}\ }\href {\doibase 10.1017/jfm.2012.418}
  {\bibfield  {journal} {\bibinfo  {journal} {J. Fluid Mech.}\ }\textbf
  {\bibinfo {volume} {712}},\ \bibinfo {pages} {273--294} (\bibinfo {year}
  {2012})}\BibitemShut {NoStop}%
\bibitem [{\citenamefont {Gat}\ and\ \citenamefont {Gharib}(2013)}]{gat2013}%
  \BibitemOpen
  \bibfield  {author} {\bibinfo {author} {\bibfnamefont {A.~D.}\ \bibnamefont
  {Gat}}\ and\ \bibinfo {author} {\bibfnamefont {M.}~\bibnamefont {Gharib}},\
  }\bibfield  {title} {\enquote {\bibinfo {title} {Elasto-capillary coalescence
  of multiple parallel sheets},}\ }\href {\doibase 10.1017/jfm.2013.86}
  {\bibfield  {journal} {\bibinfo  {journal} {J. Fluid Mech.}\ }\textbf
  {\bibinfo {volume} {723}},\ \bibinfo {pages} {692--705} (\bibinfo {year}
  {2013})}\BibitemShut {NoStop}%
\bibitem [{\citenamefont {Singh}\ \emph {et~al.}(2014)\citenamefont {Singh},
  \citenamefont {Lister},\ and\ \citenamefont {Vella}}]{singh2014}%
  \BibitemOpen
  \bibfield  {author} {\bibinfo {author} {\bibfnamefont {K.}~\bibnamefont
  {Singh}}, \bibinfo {author} {\bibfnamefont {J.~R.}\ \bibnamefont {Lister}}, \
  and\ \bibinfo {author} {\bibfnamefont {D.}~\bibnamefont {Vella}},\ }\bibfield
   {title} {\enquote {\bibinfo {title} {A fluid-mechanical model of
  elastocapillary coalescence},}\ }\href {\doibase 10.1017/jfm.2014.102}
  {\bibfield  {journal} {\bibinfo  {journal} {J. Fluid Mech.}\ }\textbf
  {\bibinfo {volume} {745}},\ \bibinfo {pages} {621--646} (\bibinfo {year}
  {2014})}\BibitemShut {NoStop}%
\bibitem [{\citenamefont {Anoop}\ and\ \citenamefont {Sen}(2015)}]{anoop2015}%
  \BibitemOpen
  \bibfield  {author} {\bibinfo {author} {\bibfnamefont {R.}~\bibnamefont
  {Anoop}}\ and\ \bibinfo {author} {\bibfnamefont {A.~K.}\ \bibnamefont
  {Sen}},\ }\bibfield  {title} {\enquote {\bibinfo {title} {Capillary flow
  enhancement in rectangular polymer microchannels with a deformable wall},}\
  }\href {\doibase 10.1103/physreve.92.013024} {\bibfield  {journal} {\bibinfo
  {journal} {Phys. Rev. E}\ }\textbf {\bibinfo {volume} {92}},\ \bibinfo
  {pages} {013024} (\bibinfo {year} {2015})}\BibitemShut {NoStop}%
\bibitem [{\citenamefont {Lee}\ \emph {et~al.}(2016)\citenamefont {Lee},
  \citenamefont {Kim}, \citenamefont {Kim},\ and\ \citenamefont
  {Mahadevan}}]{lee2016}%
  \BibitemOpen
  \bibfield  {author} {\bibinfo {author} {\bibfnamefont {M.}~\bibnamefont
  {Lee}}, \bibinfo {author} {\bibfnamefont {S.}~\bibnamefont {Kim}}, \bibinfo
  {author} {\bibfnamefont {H.}~\bibnamefont {Kim}}, \ and\ \bibinfo {author}
  {\bibfnamefont {L.}~\bibnamefont {Mahadevan}},\ }\bibfield  {title} {\enquote
  {\bibinfo {title} {Bending and buckling of wet paper},}\ }\href {\doibase
  http://dx.doi.org/10.1063/1.4944659} {\bibfield  {journal} {\bibinfo
  {journal} {Phys. Fluids}\ }\textbf {\bibinfo {volume} {28}},\ \bibinfo
  {pages} {042101} (\bibinfo {year} {2016})}\BibitemShut {NoStop}%
\bibitem [{\citenamefont {Martinez}\ \emph {et~al.}(2008)\citenamefont
  {Martinez}, \citenamefont {Phillips},\ and\ \citenamefont
  {Whitesides}}]{martinez2008}%
  \BibitemOpen
  \bibfield  {author} {\bibinfo {author} {\bibfnamefont {A.~W.}\ \bibnamefont
  {Martinez}}, \bibinfo {author} {\bibfnamefont {S.~T.}\ \bibnamefont
  {Phillips}}, \ and\ \bibinfo {author} {\bibfnamefont {G.~M.}\ \bibnamefont
  {Whitesides}},\ }\bibfield  {title} {\enquote {\bibinfo {title}
  {Three-dimensional microfluidic devices fabricated in layered paper and
  tape},}\ }\href {\doibase 10.1073/pnas.0810903105} {\bibfield  {journal}
  {\bibinfo  {journal} {Proc. Natl. Acad. Sci.}\ }\textbf {\bibinfo {volume}
  {105}},\ \bibinfo {pages} {19606--19611} (\bibinfo {year}
  {2008})}\BibitemShut {NoStop}%
\bibitem [{\citenamefont {Topgaard}\ and\ \citenamefont
  {S\"{o}derman}(2001)}]{topgaard2001}%
  \BibitemOpen
  \bibfield  {author} {\bibinfo {author} {\bibfnamefont {D.}~\bibnamefont
  {Topgaard}}\ and\ \bibinfo {author} {\bibfnamefont {O.}~\bibnamefont
  {S\"{o}derman}},\ }\bibfield  {title} {\enquote {\bibinfo {title} {Diffusion
  of water absorbed in cellulose fibers studied with {1H-NMR}},}\ }\href
  {\doibase 10.1021/la000982l} {\bibfield  {journal} {\bibinfo  {journal}
  {Langmuir}\ }\textbf {\bibinfo {volume} {17}},\ \bibinfo {pages} {2694--2702}
  (\bibinfo {year} {2001})}\BibitemShut {NoStop}%
\bibitem [{\citenamefont {Enderby}(1955)}]{enderby1955}%
  \BibitemOpen
  \bibfield  {author} {\bibinfo {author} {\bibfnamefont {J.~A.}\ \bibnamefont
  {Enderby}},\ }\bibfield  {title} {\enquote {\bibinfo {title} {Water
  absorption by polymers},}\ }\href {\doibase 10.1039/tf9555100106} {\bibfield
  {journal} {\bibinfo  {journal} {J. Chem. Soc. Faraday Trans.}\ }\textbf
  {\bibinfo {volume} {51}},\ \bibinfo {pages} {106} (\bibinfo {year}
  {1955})}\BibitemShut {NoStop}%
\bibitem [{\citenamefont {Kvick}\ \emph {et~al.}(2017)\citenamefont {Kvick},
  \citenamefont {Martinez}, \citenamefont {Hewitt},\ and\ \citenamefont
  {Balmforth}}]{kvick2017}%
  \BibitemOpen
  \bibfield  {author} {\bibinfo {author} {\bibfnamefont {M.}~\bibnamefont
  {Kvick}}, \bibinfo {author} {\bibfnamefont {D.~M.}\ \bibnamefont {Martinez}},
  \bibinfo {author} {\bibfnamefont {D.~R.}\ \bibnamefont {Hewitt}}, \ and\
  \bibinfo {author} {\bibfnamefont {N.~J.}\ \bibnamefont {Balmforth}},\
  }\bibfield  {title} {\enquote {\bibinfo {title} {Imbibition with swelling:
  Capillary rise in thin deformable porous media},}\ }\href {\doibase
  10.1103/physrevfluids.2.074001} {\bibfield  {journal} {\bibinfo  {journal}
  {Phys. Rev. Fluids}\ }\textbf {\bibinfo {volume} {2}},\ \bibinfo {pages}
  {074001} (\bibinfo {year} {2017})}\BibitemShut {NoStop}%
\bibitem [{\citenamefont {Ha}\ \emph {et~al.}(2018)\citenamefont {Ha},
  \citenamefont {Kim}, \citenamefont {Jung}, \citenamefont {Yun}, \citenamefont
  {Kim},\ and\ \citenamefont {Kim}}]{ha2018}%
  \BibitemOpen
  \bibfield  {author} {\bibinfo {author} {\bibfnamefont {J.}~\bibnamefont
  {Ha}}, \bibinfo {author} {\bibfnamefont {J.}~\bibnamefont {Kim}}, \bibinfo
  {author} {\bibfnamefont {Y.}~\bibnamefont {Jung}}, \bibinfo {author}
  {\bibfnamefont {G.}~\bibnamefont {Yun}}, \bibinfo {author} {\bibfnamefont
  {D.}~\bibnamefont {Kim}}, \ and\ \bibinfo {author} {\bibfnamefont
  {H.}~\bibnamefont {Kim}},\ }\bibfield  {title} {\enquote {\bibinfo {title}
  {Poro-elasto-capillary wicking of cellulose sponges},}\ }\href {\doibase
  10.1126/sciadv.aao7051} {\bibfield  {journal} {\bibinfo  {journal} {Sci.
  Adv.}\ }\textbf {\bibinfo {volume} {4}},\ \bibinfo {pages} {eaao7051}
  (\bibinfo {year} {2018})}\BibitemShut {NoStop}%
\bibitem [{\citenamefont {Holmes}\ \emph {et~al.}(2016)\citenamefont {Holmes},
  \citenamefont {Brun}, \citenamefont {Pandey},\ and\ \citenamefont
  {Proti{\`{e}}re}}]{holmes2016}%
  \BibitemOpen
  \bibfield  {author} {\bibinfo {author} {\bibfnamefont {D.~P.}\ \bibnamefont
  {Holmes}}, \bibinfo {author} {\bibfnamefont {P.-T.}\ \bibnamefont {Brun}},
  \bibinfo {author} {\bibfnamefont {A.}~\bibnamefont {Pandey}}, \ and\ \bibinfo
  {author} {\bibfnamefont {S.}~\bibnamefont {Proti{\`{e}}re}},\ }\bibfield
  {title} {\enquote {\bibinfo {title} {Rising beyond elastocapillarity},}\
  }\href {\doibase 10.1039/c6sm00329j} {\bibfield  {journal} {\bibinfo
  {journal} {Soft Matter}\ }\textbf {\bibinfo {volume} {12}},\ \bibinfo {pages}
  {4886--4890} (\bibinfo {year} {2016})}\BibitemShut {NoStop}%
\bibitem [{\citenamefont {Reyssat}\ and\ \citenamefont
  {Mahadevan}(2011)}]{reyssat2011}%
  \BibitemOpen
  \bibfield  {author} {\bibinfo {author} {\bibfnamefont {E.}~\bibnamefont
  {Reyssat}}\ and\ \bibinfo {author} {\bibfnamefont {L.}~\bibnamefont
  {Mahadevan}},\ }\bibfield  {title} {\enquote {\bibinfo {title} {How wet paper
  curls},}\ }\href {\doibase 10.1209/0295-5075/93/54001} {\bibfield  {journal}
  {\bibinfo  {journal} {Europhys. Lett.}\ }\textbf {\bibinfo {volume} {93}},\
  \bibinfo {pages} {54001} (\bibinfo {year} {2011})}\BibitemShut {NoStop}%
\bibitem [{\citenamefont {Capodicasa}\ \emph {et~al.}(2010)\citenamefont
  {Capodicasa}, \citenamefont {Fedi}, \citenamefont {Porcelli},\ and\
  \citenamefont {Zannoni}}]{capodicasa2010}%
  \BibitemOpen
  \bibfield  {author} {\bibinfo {author} {\bibfnamefont {Serena}\ \bibnamefont
  {Capodicasa}}, \bibinfo {author} {\bibfnamefont {Stefano}\ \bibnamefont
  {Fedi}}, \bibinfo {author} {\bibfnamefont {Anna~Maria}\ \bibnamefont
  {Porcelli}}, \ and\ \bibinfo {author} {\bibfnamefont {Davide}\ \bibnamefont
  {Zannoni}},\ }\bibfield  {title} {\enquote {\bibinfo {title} {The microbial
  community dwelling on a biodeteriorated 16th century painting},}\ }\href
  {\doibase 10.1016/j.ibiod.2010.08.006} {\bibfield  {journal} {\bibinfo
  {journal} {Int. Biodeterior. Biodegradation}\ }\textbf {\bibinfo {volume}
  {64}},\ \bibinfo {pages} {727--733} (\bibinfo {year} {2010})}\BibitemShut
  {NoStop}%
\bibitem [{\citenamefont {Martinez}\ \emph {et~al.}(2010)\citenamefont
  {Martinez}, \citenamefont {Phillips}, \citenamefont {Whitesides},\ and\
  \citenamefont {Carrilho}}]{martinez2010}%
  \BibitemOpen
  \bibfield  {author} {\bibinfo {author} {\bibfnamefont {A.~W.}\ \bibnamefont
  {Martinez}}, \bibinfo {author} {\bibfnamefont {S.~T.}\ \bibnamefont
  {Phillips}}, \bibinfo {author} {\bibfnamefont {G.~M.}\ \bibnamefont
  {Whitesides}}, \ and\ \bibinfo {author} {\bibfnamefont {E.}~\bibnamefont
  {Carrilho}},\ }\bibfield  {title} {\enquote {\bibinfo {title} {Diagnostics
  for the developing world: Microfluidic paper-based analytical devices},}\
  }\href {\doibase 10.1021/ac9013989} {\bibfield  {journal} {\bibinfo
  {journal} {Anal. Chem.}\ }\textbf {\bibinfo {volume} {82}},\ \bibinfo {pages}
  {3--10} (\bibinfo {year} {2010})}\BibitemShut {NoStop}%
\bibitem [{\citenamefont {Qu{\'{e}}r{\'{e}}}(1997)}]{quere1997}%
  \BibitemOpen
  \bibfield  {author} {\bibinfo {author} {\bibfnamefont {D.}~\bibnamefont
  {Qu{\'{e}}r{\'{e}}}},\ }\bibfield  {title} {\enquote {\bibinfo {title}
  {Inertial capillarity},}\ }\href {\doibase 10.1209/epl/i1997-00389-2}
  {\bibfield  {journal} {\bibinfo  {journal} {Europhys. Lett.}\ }\textbf
  {\bibinfo {volume} {39}},\ \bibinfo {pages} {533--538} (\bibinfo {year}
  {1997})}\BibitemShut {NoStop}%
\bibitem [{\citenamefont {Das}\ \emph {et~al.}(2012)\citenamefont {Das},
  \citenamefont {Waghmare},\ and\ \citenamefont {Mitra}}]{das2012}%
  \BibitemOpen
  \bibfield  {author} {\bibinfo {author} {\bibfnamefont {Siddhartha}\
  \bibnamefont {Das}}, \bibinfo {author} {\bibfnamefont {Prashant~R.}\
  \bibnamefont {Waghmare}}, \ and\ \bibinfo {author} {\bibfnamefont
  {Sushanta~K.}\ \bibnamefont {Mitra}},\ }\bibfield  {title} {\enquote
  {\bibinfo {title} {Early regimes of capillary filling},}\ }\href {\doibase
  10.1103/PhysRevE.86.067301} {\bibfield  {journal} {\bibinfo  {journal} {Phys.
  Rev. E}\ }\textbf {\bibinfo {volume} {86}},\ \bibinfo {pages} {067301}
  (\bibinfo {year} {2012})}\BibitemShut {NoStop}%
\bibitem [{\citenamefont {Das}\ and\ \citenamefont {Mitra}(2013)}]{das2013}%
  \BibitemOpen
  \bibfield  {author} {\bibinfo {author} {\bibfnamefont {Siddhartha}\
  \bibnamefont {Das}}\ and\ \bibinfo {author} {\bibfnamefont {Sushanta~K.}\
  \bibnamefont {Mitra}},\ }\bibfield  {title} {\enquote {\bibinfo {title}
  {Different regimes in vertical capillary filling},}\ }\href {\doibase
  10.1103/PhysRevE.87.063005} {\bibfield  {journal} {\bibinfo  {journal} {Phys.
  Rev. E}\ }\textbf {\bibinfo {volume} {87}},\ \bibinfo {pages} {063005}
  (\bibinfo {year} {2013})}\BibitemShut {NoStop}%
\bibitem [{\citenamefont {de~Gennes}\ \emph {et~al.}(2004)\citenamefont
  {de~Gennes}, \citenamefont {Brochard-Wyart},\ and\ \citenamefont
  {Qu{\'{e}}r{\'{e}}}}]{degennes2004}%
  \BibitemOpen
  \bibfield  {author} {\bibinfo {author} {\bibfnamefont {P.}~\bibnamefont
  {de~Gennes}}, \bibinfo {author} {\bibfnamefont {F.}~\bibnamefont
  {Brochard-Wyart}}, \ and\ \bibinfo {author} {\bibfnamefont {D.}~\bibnamefont
  {Qu{\'{e}}r{\'{e}}}},\ }\href {\doibase 10.1007/978-0-387-21656-0} {\emph
  {\bibinfo {title} {Capillarity and Wetting Phenomena}}}\ (\bibinfo
  {publisher} {Springer New York},\ \bibinfo {year} {2004})\BibitemShut
  {NoStop}%
\bibitem [{\citenamefont {Nguyen}\ \emph {et~al.}(2013)\citenamefont {Nguyen},
  \citenamefont {Feng}, \citenamefont {Le}, \citenamefont {Le}, \citenamefont
  {Hoang}, \citenamefont {Tan},\ and\ \citenamefont {Duong}}]{nguyen2013}%
  \BibitemOpen
  \bibfield  {author} {\bibinfo {author} {\bibfnamefont {S.~T.}\ \bibnamefont
  {Nguyen}}, \bibinfo {author} {\bibfnamefont {J.}~\bibnamefont {Feng}},
  \bibinfo {author} {\bibfnamefont {N.~T.}\ \bibnamefont {Le}}, \bibinfo
  {author} {\bibfnamefont {A.~T.~T.}\ \bibnamefont {Le}}, \bibinfo {author}
  {\bibfnamefont {N.}~\bibnamefont {Hoang}}, \bibinfo {author} {\bibfnamefont
  {V.~B.~C.}\ \bibnamefont {Tan}}, \ and\ \bibinfo {author} {\bibfnamefont
  {H.~M.}\ \bibnamefont {Duong}},\ }\bibfield  {title} {\enquote {\bibinfo
  {title} {Cellulose aerogel from paper waste for crude oil spill cleaning},}\
  }\href {\doibase 10.1021/ie4032567} {\bibfield  {journal} {\bibinfo
  {journal} {Ind. Eng. Chem. Res.}\ }\textbf {\bibinfo {volume} {52}},\
  \bibinfo {pages} {18386--18391} (\bibinfo {year} {2013})}\BibitemShut
  {NoStop}%
\bibitem [{\citenamefont {Gui}\ \emph {et~al.}(2013)\citenamefont {Gui},
  \citenamefont {Zeng}, \citenamefont {Lin}, \citenamefont {Gan}, \citenamefont
  {Xiang}, \citenamefont {Zhu}, \citenamefont {Cao},\ and\ \citenamefont
  {Tang}}]{gui2013}%
  \BibitemOpen
  \bibfield  {author} {\bibinfo {author} {\bibfnamefont {X.}~\bibnamefont
  {Gui}}, \bibinfo {author} {\bibfnamefont {Z.}~\bibnamefont {Zeng}}, \bibinfo
  {author} {\bibfnamefont {Z.}~\bibnamefont {Lin}}, \bibinfo {author}
  {\bibfnamefont {Q.}~\bibnamefont {Gan}}, \bibinfo {author} {\bibfnamefont
  {R.}~\bibnamefont {Xiang}}, \bibinfo {author} {\bibfnamefont
  {Y.}~\bibnamefont {Zhu}}, \bibinfo {author} {\bibfnamefont {A.}~\bibnamefont
  {Cao}}, \ and\ \bibinfo {author} {\bibfnamefont {Z.}~\bibnamefont {Tang}},\
  }\bibfield  {title} {\enquote {\bibinfo {title} {Magnetic and highly
  recyclable macroporous carbon nanotubes for spilled oil sorption and
  separation},}\ }\href {\doibase 10.1021/am4015007} {\bibfield  {journal}
  {\bibinfo  {journal} {{ACS} Appl. Mater. Interfaces}\ }\textbf {\bibinfo
  {volume} {5}},\ \bibinfo {pages} {5845--5850} (\bibinfo {year}
  {2013})}\BibitemShut {NoStop}%
\end{thebibliography}%

\end{document}